\documentclass[useAMS,usenatbib]{mn2e}

\title[The star formation rate and
 surface density]
{A simple model for the relationship between star formation and
surface density}
\author[C. L. Dobbs \& J. E. Pringle]
{C. L. Dobbs\thanks{E-mail:
dobbs@astro.ex.ac.uk}$^1$ \& J. E. Pringle$^2$ \\
$^1$ School of Physics, University of Exeter, 
Stocker Road, Exeter, EX4 4QL \\
$^2$ Institute of Astronomy, Madingley Road, Cambridge, CB3 0HA \\}
\usepackage{amssymb}
\usepackage{amsmath}
\usepackage{graphicx}
\usepackage{epsfig}

\begin{document}
\date{\today}

\pagerange{\pageref{firstpage}--\pageref{lastpage}} \pubyear{0000}

\maketitle

\label{firstpage}
 \begin{abstract}
   We investigate the relationship between the star formation rate per
   unit area and the surface density of the ISM (the local
   Kennicutt-Schmitt law) using a simplified model of the ISM and a
   simple estimate of the star formation rate based on the mass of gas
   in bound clumps, the local dynamical timescales of the clumps, and
   an efficiency parameter of around $\epsilon \approx 5$ per
   cent. Despite the simplicity of the approach, we are able to
   reproduce the observed linear relation between star formation rate
   and surface density of dense (molecular) gas. We use a simple model
   for the dependence of H$_2$ fraction on total surface density to
   argue why neither total surface density nor the H\textsc{i} surface
   density are good local indicators of star formation rate. We also
   investigate the dependence of the star formation rate on the depth
   of the spiral potential. Our model indicates that the mean star
   formation rate does not depend significantly on the strength of the
   spiral potential, but that a stronger spiral potential, for a given
   mean surface density, does result in more of the star formation
   occurring close to the spiral arms.  This agrees with the
   observation that grand design galaxies do not appear to show a
   larger degree of star formation compared to their flocculent
   counterparts.
 \end{abstract}

\begin{keywords}
  stars: formation -- galaxies: spiral --  galaxies: kinematic and
  dynamics -- MHD -- ISM: clouds -- ISM: evolution
\end{keywords}

\section{Introduction}

\subsection{Observational Background}

The dependence of the star formation rate per unit area on the surface
density of the ISM in a galaxy is one of the most highly speculated
problems in extra-galactic surveys, as well as theoretical analysis of
star formation. The star formation law predicts both the amount of
star formation and the degree of stellar feedback in a galaxy, and is
consequently a vital requirement for models of galaxy evolution
(e.g. \citealt*{Tan1999,vBosch2000,Springel2000,Hou2000,Springel2003}).

\citet{Schmidt1959} related the star formation rate per unit volume,
$\rho_{\rm SFR}$, to the gas density, $\rho$, in a galaxy according to
$\rho_{\rm SFR} \propto \rho^m$. An observational law for star
formation rate per unit surface area, $\Sigma_{\rm SFR}$ in terms of
the mean gas galactic surface density, $\Sigma$, viz.  $\Sigma_{\rm
  SFR} \propto \Sigma^n$ was established by \citet{Kennicutt1989}, and
since then numerous surveys have attempted to find a universal value
of $n$. \citet{Kennicutt1989,Kennicutt1998} investigated a global star
formation law in star-forming galaxies and found $n=1.3\pm0.3$ from a
sample of 97 galaxies including normal spirals and starbursts, for gas
surface densities of a few M$_{\odot}$\,pc$^{-2}$ to around 
$10^4$\,M$_{\odot}$\,pc$^{-2}$, although it should be noted that at a given
$\Sigma$, the scatter in $\log \Sigma_{\rm SFR}$ is generally more
than $\pm 0.5$.  Other surveys include \citet{Gao2004}, who
trace both the densest gas, using HCN, and the total gas in a
galaxy. They obtain different relations depending on what is meant by
$\Sigma$. For $\Sigma$ measuring dense gas they find $n$=1 whereas for
$\Sigma$ measuring total gas density they find $1.73$.
  
The local relationship between star formation rate per unit area and
surface density (what we call here the {\em local} Kennicutt-Schmidt
law), has also been examined for individual galaxies. \citet{Wong2002}
used radially averaged values of the star formation rate and surface
densities for 6 galaxies, finding n=1.2--2.1.  Kennicutt et. al. 2007
also obtained $n=1.56\pm0.04$ for M51, by sampling the H$_{\alpha}$
emission and total surface density over 500\,pc size regions (see also
\citealt{Schuster2007}). There is a slightly less steep correlation
($n=1.37\pm0.03)$ for the surface density of molecular gas, but no
correlation with the HI gas. Similarly \citet{Heyer2004} found
$n=1.36$ for M33, when considering the molecular gas alone, but a much
steeper dependence ($n=3.3\pm0.1$) on the total gas surface density.

In fact the general consensus from these results is that a
well-defined local star formation law holds only for the molecular gas
in these galaxies. This conclusion is further endorsed by recent
results from the THINGS survey \citep{Bigiel2008}, which show that
there is no correlation between the H\textsc{i} and the star formation
rate -- $\Sigma_{\rm SFR}$ is multivalued for a given density of
H\textsc{i}. Their results also indicate a fairly sharp transition
from a regime where gas is predominantly H\textsc{i}, to where gas is
mainly molecular. Thus in the H\textsc{i} regime, $\Sigma_{\rm SFR}$
varies very steeply with $\Sigma$, whereas in the H$_2$ regime, $n$ is
roughly 1. For the total gas surface density, $n$ lies between 1 and
3, although it is evident that the data does not give a good linear
fit.

\subsection{Theoretical interpretations}
\label{introtheory}

Several theoretical explanations of the Kennicutt-Schmidt law have
also been advanced. All interpret $\Sigma$ as being the total surface
gas density. Since star formation occurs in a chaotic or turbulent
environment, a natural explanation is that turbulence somehow
regulates how much gas exceeds the high densities required for star
formation. \citet{Elmegreen2002} and \citet{Krum2005} assume a probability density function of densities in a turbulent regime to obtain a star formation law with $n=1.4$. For \citet{Elmegreen2002}, this involves assuming an unknown star formation efficiency ($\epsilon$), but \citet{Krum2005} instead determine the star formation rate efficiency, thus eliminating $\epsilon$. This is essentially the derived star formation rate divided by the total possible star formation rate if all gas at a particular density was turned into stars.

\citet{Krum2007a} and \citet{Krum2007b} further compare the turbulence regulating model of star formation with observations. \citet{Krum2005} predict that the star formation rate efficiency is $\sim 0.013$, independent of density. Their observed estimates of the star formation rate efficiency \citep{Krum2007a} are consistent with this prediction. However the uncertainties (in particular for lifetimes for a given tracer) do not rule out a star formation rate efficiency which increases with density either. \citet{Krum2007b} also determine CO and HCN luminosities, and show that the star formation rate is linear with relation to L(HCN) and nonlinear for L(CO).

An alternative possibility is to
estimate the star formation rate from cloud-cloud collisions. In this
case $\Sigma_{\rm SFR}$ is proportional to the surface density over
the collision time, which in turn is dependent on the shear in the
disc, measured by the orbital angular velocity $\Omega$. This yields
an alternative form of the Schmidt law, $\Sigma_{\rm SFR} \propto
\epsilon \Omega \Sigma^N$ where $\epsilon$ is the star formation
efficiency and $N=1$ \citep{Wyse1986,Wyse1989,Tan2000}. \citet{Silk1997}
explicitly includes the star formation efficiency by assuming that
star formation is self regulating in the disc, obtaining $\Sigma_{SFR}
\propto \Sigma^{1.78}$.

In addition to the observational surveys, several numerical
simulations have investigated star formation rates in galaxy simulations
\citep{Li2006,Robertson2008,Tasker2008b}. 
However a major problem with
numerical evaluations of the star formation rate is that it is
dependent on what assumptions are made about the conditions for
forming sink particles at high densities, or for taking account of
possible forms of stellar feedback. A density threshold is required,
as well as a star formation efficiency, and so the degree of star
formation in these simulations is also dependent on both these
parameters. \citet{Saitoh2008} performed simulations with varying
thresholds and star formation efficiencies, concluding that the star
formation rates and Kennicutt-Schmidt law do not significantly change
provided there is a high threshold density. However an estimation of
the star formation rates from first principles, without these
approximations, is beyond current capabilities.

\citet{Tasker2008c} perform similar models to those of the current paper, and likewise do not include any star formation prescription. In their simulations there is no underlying spiral potential, rather numerous flocculent spiral arms cover the disc. In agreement with \citet{Dobbs2008} they find that GMCs form by agglomeration via collisions, although collisions occur throughout the disc whereas in \citet{Dobbs2008}, collisions between clouds are largely confined to the spiral arms. They do not make any estimates of the star formation rate from their results, however they do show that the number of collisions between clouds is consistent with the model presented in \citet{Tan2000}.

\subsection{The current paper}
One approach to trying to understand the physics underlying the basis
of the Schmitt-Kennicutt relations is to undertake ever more detailed
numerical simulations with ever increasing quantities of input physics
(e.g. \citealt{Susa2008, Shetty2008, Agertz2009}).  In
this paper, we take an alternative approach and ask the question: how
little input physics, and how few assumptions do we have to make, in
order to obtain relations which resemble the observational findings to
a reasonable degree? In this manner we hope to be able to obtain some
understanding of what the fundamental drivers for such a relationship
might be.

The calculations presented in this paper make use of the simplified
numerical simulations which have already been published in
\citet{Dobbs2008}. The simulations were performed using Smoothed
Particle Hydrodynamics, and include magnetic fields and self-gravity.
We do not include the star formation process itself nor any subsequent
stellar feedback. Thus the gas is supported against collapse by
thermal and magnetic pressure in the lower surface density
calculations. In the higher surface density results, we run the
calculations until collapse occurs. The difficulty with the assumed
lack of feedback is that we are restricted to relatively low surface
density calculations. The calculations which have been selected for
analysis in this paper are shown in Table~\ref{runs}. These calculations, with the exception of model C, are described in \citet{Dobbs2008}, but we also provide details below.

\subsubsection{Details of numerical models}
The calculations model a 3D gaseous disc between radii of 5 and 10\,kpc. The gas is assumed to orbit in a fixed galactic
gravitational potential. The potential includes a halo \citep{Caldwell1981}, disc \citep{Binney} and 4 armed spiral component \citep{Cox2002}. The gas is initially assigned velocities according to a flat rotation curve determined by the disc potential, and in addition a velocity of dispersion of 6\,km\,s$^{-1}$ is superposed.

Models A-D have
identical potentials but vary in their mean gas surface densities and
therefore compare surface density. The total mass of the disc is $1\times10^9$ M$_{\odot}$ in the 4\,M$_{\odot}$\,pc$^{-2}$ model, and 5$\times 10^9$\,M$_{\odot}$ in the 20 M$_{\odot}$\,pc$^{-2}$  model. The simulations all use 4 million particles, hence the highest mass resolution is 250\,M$_{\odot}$, and the lowest 1250\,M$_{\odot}$. 
Models A, E, F and G have the same
mean gas surface densities but adopt different strength for the spiral
potential.

In keeping with our aim for minimal input physics, the simulations are
very simplistic. In particular, all the calculations in
Table~\ref{runs} assume an interstellar medium which has two
isothermal components, one cool and one warm. We omit thermal
considerations and so there is no transition between the two phases;
the cool gas remains cool and the warm gas remains warm,
throughout. We use the same thermal distribution in all the
calculations and only vary the global surface density and/or the shock
strength. The cool gas is taken to have a temperature of $T = 100$ K,
and we will think of it as representing molecular gas (H$_2$). The
warm gas is taken to have $T = 10^4$ K, and we will think of it
representing atomic gas (H\textsc{i}). In all cases the cool and warm
gas comprise equal mass in the simulations. 
It is of course possible
to include thermal effects such as heating and cooling -- see, for
example \citep{DGCK2008} (although these did not include self-gravity)
but that is not the purpose of the current exercise. 

The initial scale heights of the warm and cold warm components are 150 and 400\,pc respectively, giving a mean smoothing length of 40\,pc. However with time the scale heights decrease to 20-100\,pc and 300\,pc.

These calculations do not include sink particles, although the gas is
self-gravitating. In the lower surface density results ($\Sigma \leq
10$\,M$_\odot$\.pc$^{-2}$), the cool gas is sufficiently supported by
magnetic and thermal pressure, and the (typically supersonic) velocity
dispersion of the gas. However in the higher surface density results,
runaway gravitational collapse does occur.  The calculations were run
for either 250\,Myr, or until the calculation is halted by
collapse. The maximum timestep in the calculations (and frequency of dumps) is 2\,Myr, but individual particles timesteps can be much less \citep{Bate1995}.

\begin{table}
\centering
\begin{tabular}{c|c|c|c|c|c|c}
 \hline 
 Model & $\Sigma$ (M$_{\odot}$\,pc$^{-2}$) & F (\%)  & Q$_c$ & Q$_h$ \\
 \hline
A &  4 & 4 & 0.5 & 5  \\
B &  8 & 4 & 0.25 & 2.5 \\
C & 16 & 4 & 0.25 & 2.5\\
D & 20 & 4 & 0.1 & 1\\
E &  4 & 2 & 0.5 & 5  \\
F & 4 & 8 & 0.5 & 5  \\
G & 4 & 16 & 0.5 & 5  \\
\hline
\end{tabular}
\caption{The calculations from \citet{Dobbs2008} which are used in
  this paper. All the calculations use 4 million particles and are
  isothermal. In these calculations, half the gas cold (100 K) and
  half warm ($10^4$ K), and the ratio of thermal to magnetic
  energy ($\beta$) is 0.4 in the cold gas. $F$ is a measure of the strength of the potential, and $Q$
  is the Toomre parameter ($Q_c$ for the cool gas and $Q_h$ for the
  warm gas). The determination of all these quantities is described in
  \citet{Dobbs2008}. Models A-D compare surface density, and E-G shock
  strength.}
\label{runs}
\end{table}

\section{Estimating $\Sigma_{\rm SFR}$ in the numerical simulations}

In this paper we shall assume that star formation occurs only in those
regions of the ISM which are bound, and at a rate determined simply by
the local dynamical timescale.

\subsection{Gravitationally bound gas}

We must first find a means of identifying which regions of the ISM are
gravitationally bound at a particular moment. It needs to be borne in
mind that while unbound gas can obviously become bound, it is also
possible for bound gas to become unbound, despite the lack of
feedback. This can come about for example as the gas accelerates out
of a spiral arm, flowing over the ridge of the spiral potential,
becoming sheared and longitudinally stretched as it does so.

We determine the mass of bound gas using the output from a simulation
at a given time frame (after 250 Myr except for models C and D, where
the times are 200 Myr and 140 Myr). The particular timeframe selected
is not important, provided the time is not too near the beginning of
the simulation, where the amount of bound gas will be underestimated.

In order to determine the mass of bound gas, we first sort the
particles according to density. We select the most dense particle, and
all particles within a radius of a smoothing length of that
particle. Then we determine the gravitational, kinetic, thermal and
magnetic energies for this group of particles. If $(E_k+2 E_{th}
+E_{mag})/E_{grav}<1$, the gas is assumed to be bound and the radius
increased until the gas becomes unbound. Then the particles which are
bound are recorded and their mass added to the total mass of bound
gas. If on the other hand the gas is unbound, the gas particles are
discarded from the list. We also performed this analysis using solely
either just the kinetic energy or both the kinetic and thermal energy.

\begin{figure}
\centerline{
\includegraphics[bb=0 20 420 390,scale=0.6]{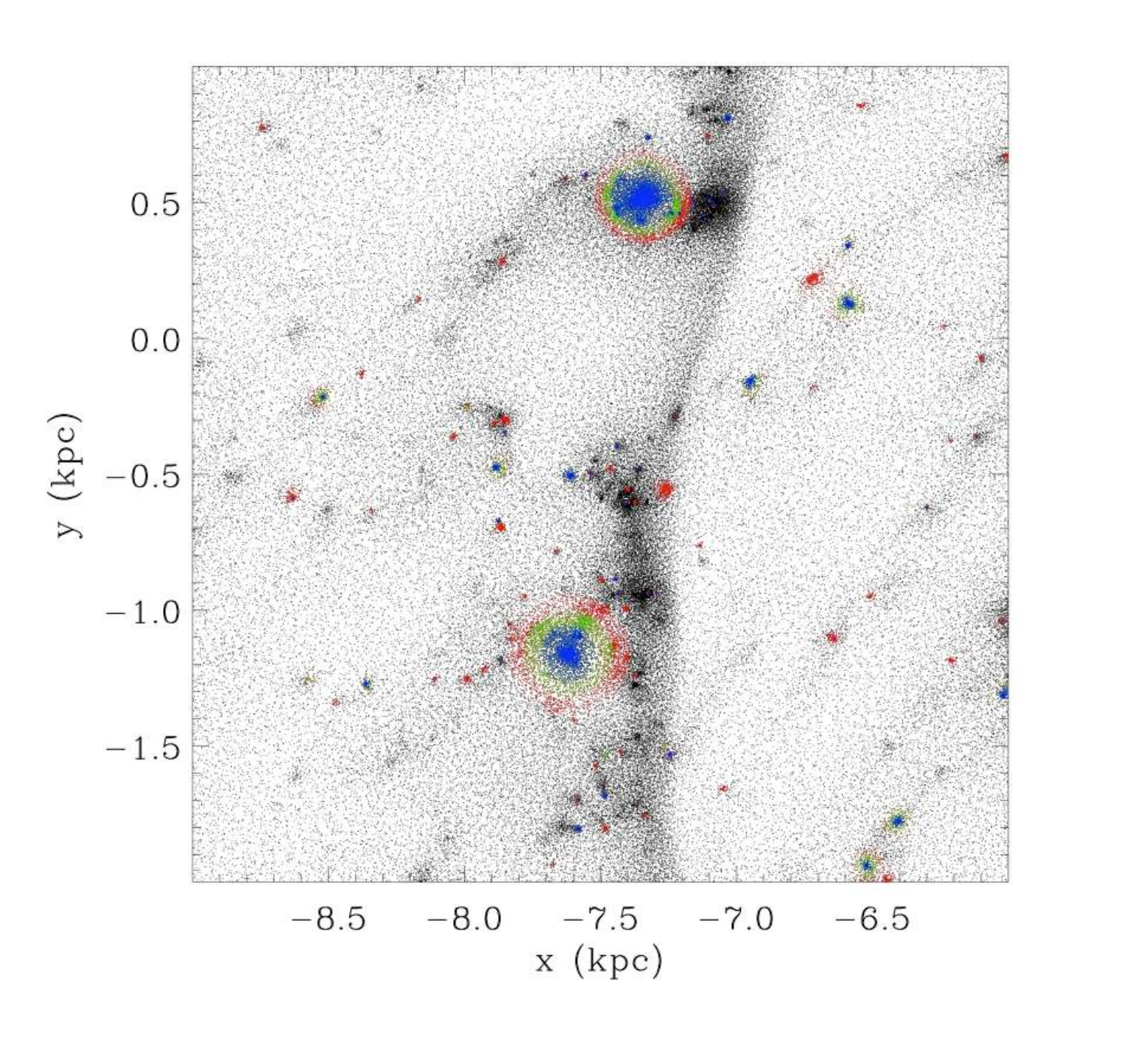}}
\caption{The location of bound gas is shown for a small section of the
  disc from the $\Sigma=8$\,M$_{\odot}$\,pc$^{-2}$ simulation Model
  B. Blue indicates regions which are bound when the kinetic, thermal
  and magnetic energies are included, green, the kinetic and thermal,
  and red, just kinetic. If only the kinetic energy is used to
  determine whether regions are bound, the clumps are clearly more
  extended. Two large and massive complexes are moving away from the
  spiral arms, which have high density and low velocity dispersions,
  so are strongly bound. There is also a considerable number of bound
  clumps in the interarm regions.}
\label{number1}
\end{figure}

Using this method, the bound gas is situated in discrete clumps (which
are spherical by assumption). Fig~\ref{number1} shows the location of
bound gas for a small section of the disc, for the simulation Model~B
with $\Sigma=8$\,M$_{\odot}$\,pc$^{-2}$. The colours indicate whether
just the kinetic, the kinetic and thermal, or the kinetic, thermal and
magnetic energies are included. As expected, when all three energies
are included, the extent of a clump which is bound becomes smaller,
whereas the clumps are more extended (particularly true in the higher
surface density calculations) when only the kinetic energy is
considered. In other simulations which investigate the
Kennicutt-Schmidt law, magnetic fields are not included
\citep*{Bottema2003,Tasker2006,Li2006,Booth2007}, so by default only
the kinetic and thermal energies can be used.

In Fig.~\ref{number1} the bound gas is dominated by two massive clumps
which are just leaving the arms.  These are more extended since they
have higher densities and lower velocity dispersions than clumps in
the spiral arms. The gravitational energy of these clumps is also high
since they are centrally condensed. Essentially these might be taken
to represent massive Giant Molecular Clouds (GMCs), which formed in
the spiral arms and are now entering the interarm regions, leaving
only less massive clumps in the spiral arms. We also see that there
are numerous bound clumps in the interarm regions. Overall in the
bound gas there is no particular evidence of spiral structure other
than the two massive `GMCs'. The degree of bound gas in the interarm
regions reflects the fact that the velocity dispersions of clumps in
the interarm regions are lower than in the spiral arms. In the
interarm regions there are fewer collisions and fewer interactions
between clumps.

\begin{figure}
\centerline{
\includegraphics[bb=150 0 420 420, scale=0.42]{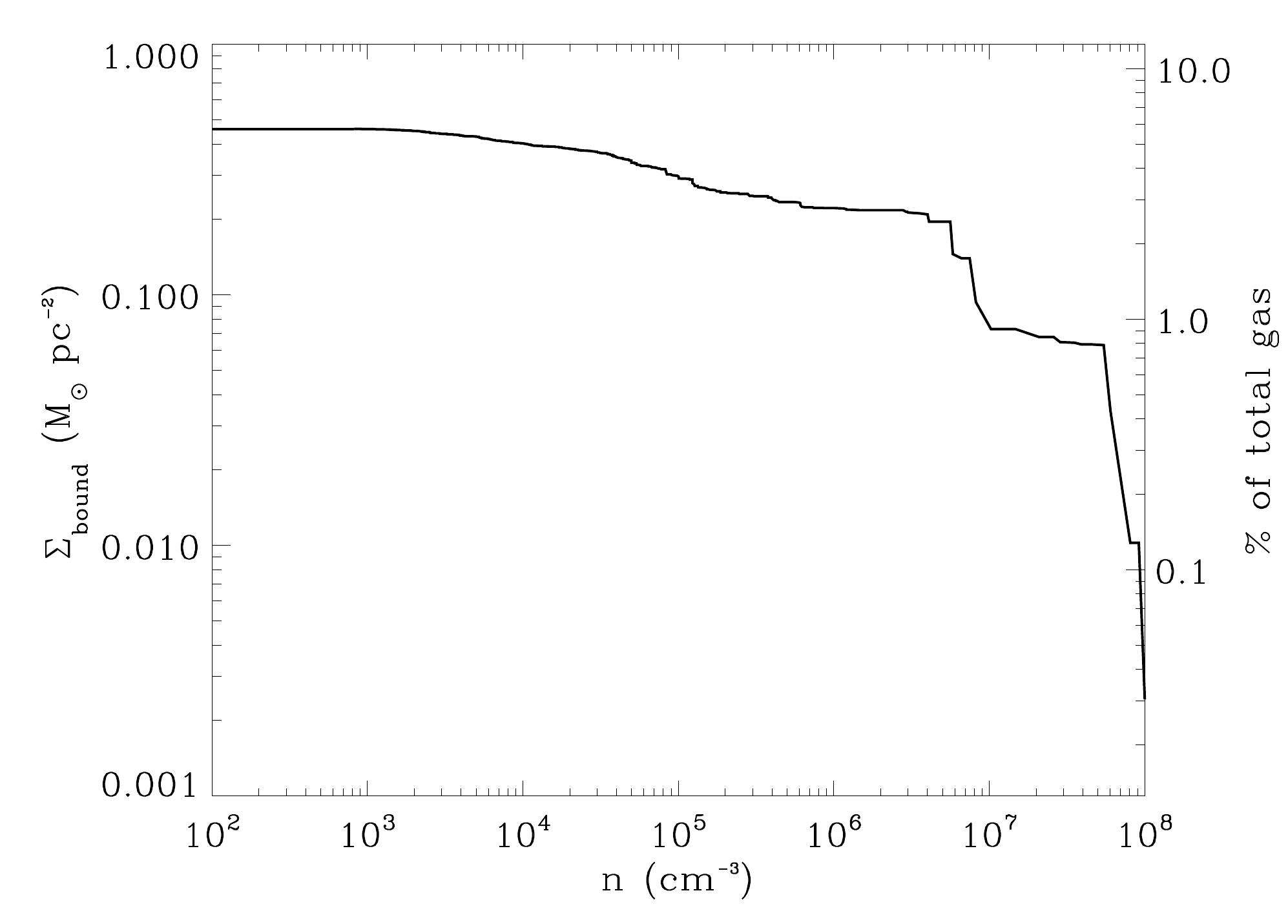}}
\caption{The mass of bound gas is plotted versus number density $n$
  when $\Sigma=8$\,M$_{\odot}$\,pc$^{-2}$ (Model B). Particles are
  sorted by density, and so $n$ represents the maximum density in a
  given bound region. The mass indicated on the $y$-axis is
  cumulative, and thus represents the total mass of bound gas in
  regions with a maximum density $>n$. The mass of bound gas is
  calculated using the kinetic, thermal and magnetic energy. Clumps
  with densities $<10^4$\,cm$^{-3}$ do not tend to contain bound
  gas. The right hand $y$-axis indicates the percentage of the total
  gas in the disc which is bound.}
\label{number2}
\end{figure}

From hereon, we shall generally assume that the calculation of bound
gas includes the thermal and magnetic energy, as well as the
kinetic. This assumption minimises the amount of 'bound' gas present
in the simulations. Using the kinetic energy alone tends to produce an
unrealistically high fraction of bound gas. In Fig.~\ref{number2} we
plot the cumulative surface density of bound gas versus number density
for model B.  Here the cumulative mean surface density of bound gas is
defined as $\Sigma_{\rm bound}(n)=M_{\rm bound}(n)/A$, where $M_{\rm
  bound}(n)$ is the mass of bound gas in clumps with maximum density
$>n$, and $A$ is the area of the galactic disc. It can be seen that
the maximum density in a clump typically must be $n>10^4$ cm$^{-3}$ to
obtain bound regions. This condition is less stringent if thermal
and/or magnetic energies are ignored. Fig.~\ref{number2} also shows
that the bound gas represents 6 \% of all the gas in the disc. 

\subsection{Star formation rate}
\label{SFRestimate}

In order to determine the star formation rate, having determined an
estimate of the amount of bound gas, we require an estimate of the
local timescale for star formation.  As we mentioned above we take
this to be the local dynamical timescale of the gas,
\begin{equation}
\label{dyntime}
  t_{\rm dyn}=\sqrt{\frac{3 \pi}{16 G \langle \rho \rangle}} \,.
\end{equation}
Here $\langle \rho \rangle$ is the {\em median} density of a
clump. An alternative would be to use the volume average density,
$\bar{\rho}$, but we find that this does not lead to a noticeable
difference in the results.  The star formation rate of a particular
clump is then assumed to be $\dot{M}=M_{\rm bound}/t_{\rm dyn}$ where
$M_{\rm bound}$ is the mass of bound gas in a clump and $t_{\rm dyn}$
the dynamical timescale of that clump given by
equation~\ref{dyntime}. The star formation rate per unit area, averaged
over a given region, is then
\begin{equation}
\label{SFR}
\Sigma_{\rm SFR}=\frac{\sum_{i=1}^{N} \dot{M_{i}}}{A}
\end{equation}
where the summation is over all ($N$) bound clumps, and $A$ is the area
of the region under consideration.

It is well known that star formation is a relatively inefficient
process in that not all the bound gas is converted to stars on a local
dynamical timescale. We can allow for this by including an efficiency
parameter $\epsilon$. In this case the star formation rate would be
\begin{equation}
\Sigma_{\rm SFR}=\frac{\sum_{i=1}^{N} \epsilon \, \dot{M_{i}}}{A}
\end{equation}
where $\epsilon$ is the constant star formation efficiency. However we
do not consider the evolution of the gas once it becomes bound. Rather
than assume a value of $\epsilon$ for the star formation rate, we
consider what value of $\epsilon$ would be required to fit
observations. As we expect the value of $\epsilon$ turns out to be
small, justifying {\em a posteriori} our decision not to remove mass
from the ISM in order to model star formation.

\section{Results}

We now apply these simple assumptions to the numerical simulations.

\subsection{Global star formation rate versus mean surface density}
\begin{figure}
\centerline{
\includegraphics[bb=80 350 600 770,scale=0.45]{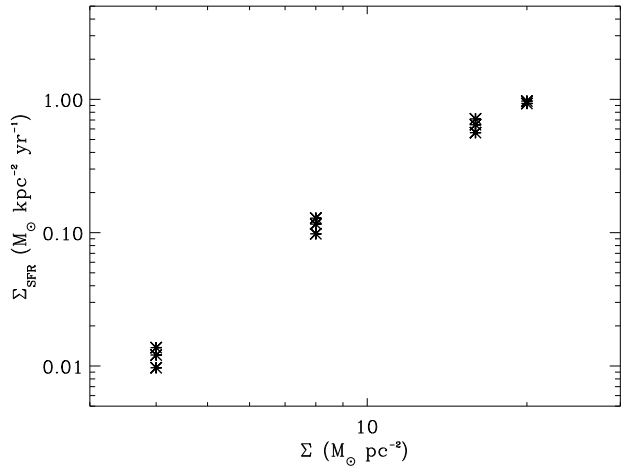}}
\caption{The estimated global star formation rate ($\Sigma_{\rm
    bound}/t_{\rm dyn}$) is shown versus mean surface density for the
  galaxies in Models A, B, C and D. The star formation rate does not
  show a good linear fit with surface density, and at low densities at
  least, is steeper than the observed Kennicutt-Schmidt law.}
\label{number3}
\end{figure}

In Fig.~\ref{number3}, we plot the star formation rate per unit area
averaged over for the whole disc in Models A-D, estimated by using
Eqn~\ref{SFR}.  This is then the equivalent of the global
Kennicutt-Schmidt law. Points are shown at three different time frames
for each surface density to illustrate the point that the ISM is not
evolving significantly through the simulations.  The points do show a
correlation of star formation rate with surface density, roughly of the form $\Sigma_{\rm SFR} \propto \Sigma ^{2.8}$.  However, even though the range in surface densities is not
large, the figure does indicate that the relation we find is not a
simple power law, with the dependence becoming shallower at higher
densities, or equivalently showing a sharp downturn at low surface densities.

The estimated star formation rates in our models are much higher than
observed since the formula in equation~\ref{SFR} assumes a star
formation efficiency of 100\%.  We can however estimate what
efficiency would correspond to observations. Comparing the star
formation rate at a surface density of 10\,M$_{\odot}$\,pc$^{-2}$ to
Fig.~2 of \citet{Kennicutt1998}, we find that we need to assume an
efficiency of 5\% ($\epsilon = 0.05$) in order to produce star
formation rates compatible with observations.

We can also consider the star formation rate efficiency by calculating the ratio of $\Sigma_{\rm bound}/\Sigma_{\rm total}$, as shown on Fig.~2. This fraction is 1.5 \%, 6 \%, 5 \% and 6 \% with surface densities of 4, 8, 16 and 20\,M$_{\odot}$\,pc$^{-2}$. Thus the proportion of gas undergoing star formation is similar in the regime where the slope of $\Sigma_{\rm sfr}$ starts to depend linearly with $\Sigma$, but somewhat lower at the lowest surface density.

\subsection{Local star formation rate versus local surface density}

\begin{figure}
\centerline{
\includegraphics[bb=0 0 420 400,scale=0.63]{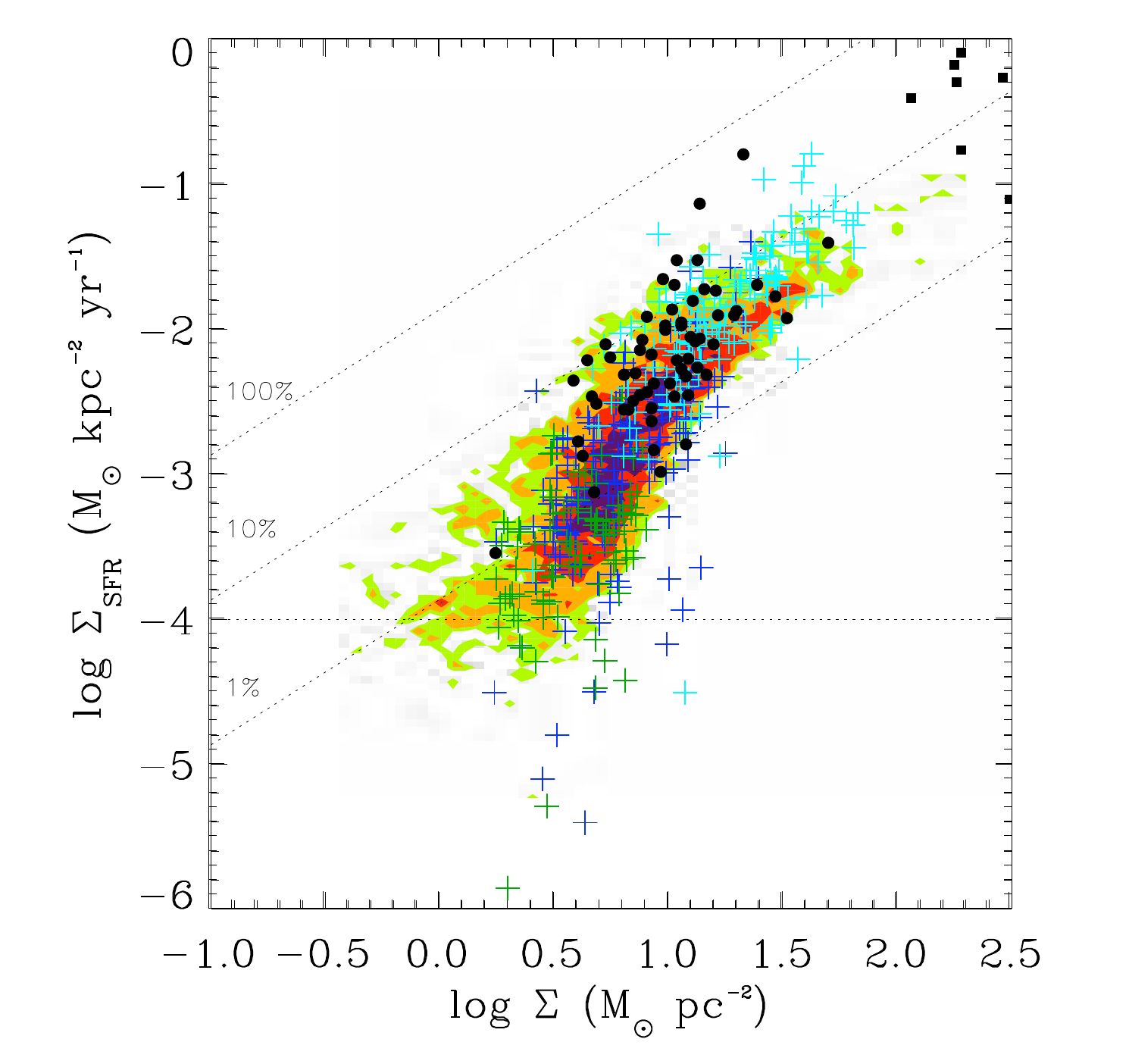}}
\caption{The star formation rate is plotted for the 4 (dark green crosses, Model
  A), 8 (blue crosses, Model B) and 20\,M$_{\odot}$\,pc$^{-2}$ (cyan crosses, Model D)
  models, assuming a star formation efficiency of  $\epsilon = 0.05$. 
  The black circles are normal spirals, and the black squares starburst galaxies from the survey of \citet{Kennicutt1989}. These are all overplotted on Fig.~8 (middle right panel) of \citet{Bigiel2008}, which is the green and orange contour plot visible in the background. This shows the star formation rate versus surface density of H\textsc{i} and H$_2$ sampled over 500\,pc$^2$ regions from 7 spiral galaxies. The green, orange, red and magenta contours indicate 1, 2, 5 and 10 data points respectively. The star formation rates from the simulations and the observed data agree reasonably, and both show a downturn in slope at around $\Sigma=10$\,M$_{\odot}$\,pc$^{-2}$.}
\label{Bigiel}
\end{figure}

With effectively only 4 data points, evaluating the dependence of the
star formation rate per unit area on surface densities averaged over
the entire disc is limited. We have therefore divided each galaxy into
500 x 500\,pc squares (the resolution of the recent THINGS results). We
then calculate the star formation rate per unit area from the surface
density of bound gas in each square, also assuming a star formation
efficiency of $\epsilon = 0.05$, in order to compare with
observations. In Fig.~\ref{Bigiel}, we have combined the star
formation rates from a quarter of the disc in the calculations of
Models A, B and D, which have mean surface densities of 4, 8, and 
20\,M$_{\odot}$\,pc$^{-2}$ into a single plot, thus acquiring more data
points and a greater range of densities. In the figure we plot for
each square the estimated star formation rate per unit area versus the
mean ISM surface density (cool and warm gas).

\begin{figure}
\centerline{
\includegraphics[scale=0.46]{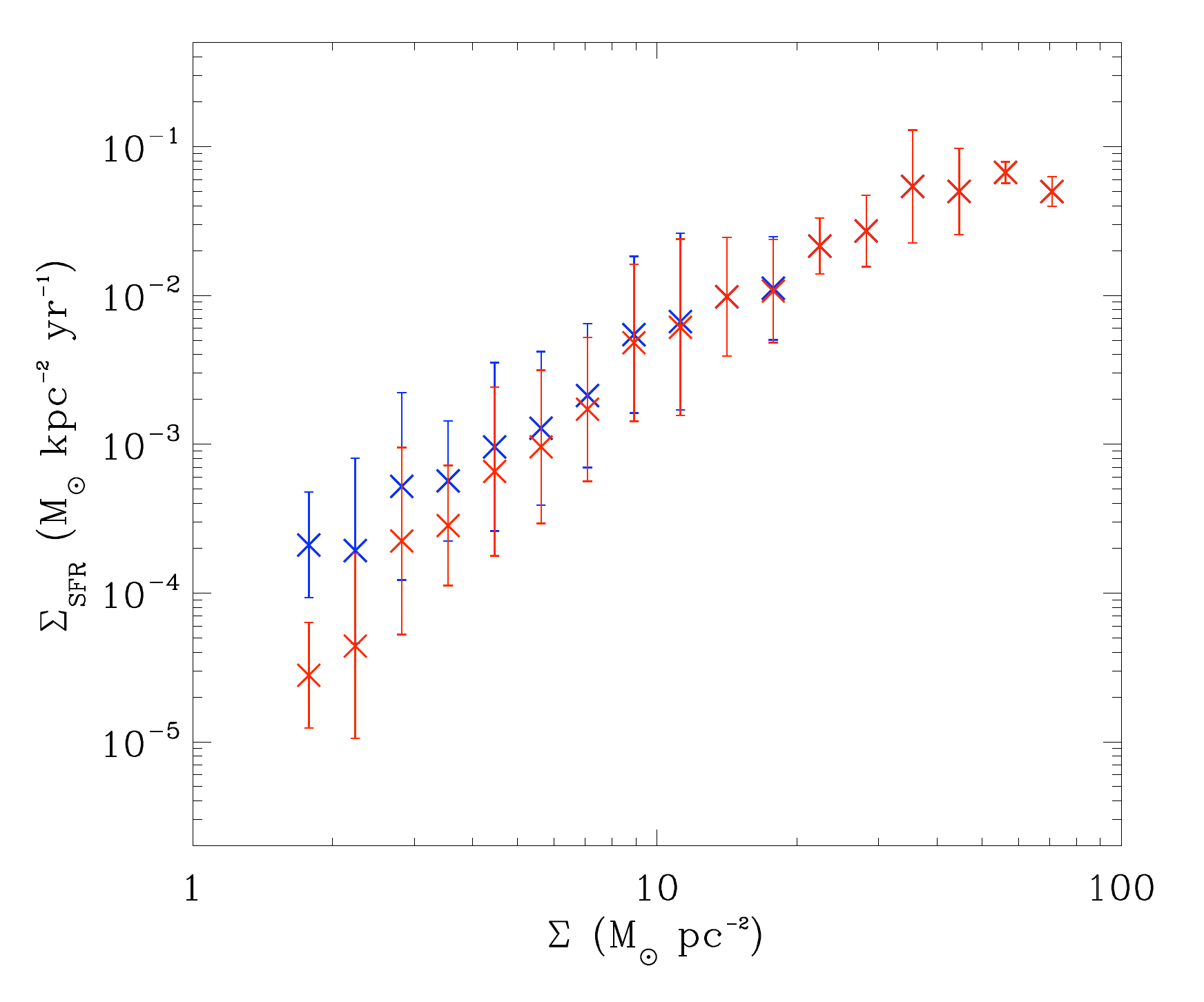}}
\caption{The star formation rate for points in the simulations with 4,  
  8 and 20\,M$_{\odot}$\,pc$^{-2}$ (as shown in Fig.~4) is binned
  according to surface density. The scatter in the average star
  formation rate is then shown as 1-$\sigma$ error bars. 
  The different points indicate whether regions without bound gas are included.
  The blue data points exclude regions in which none of the gas is bound. The red data points are
  inclusive of those regions which contain no bound gas (and are therefore not seen on Fig.~4), where instead we assume a 
  maximum star formation rate of $10^{-6}$\,M$_{\odot}$\,kpc$^{-2}$ yr$^{-1}$.}
\label{number5}
\end{figure}

In Fig.~\ref{Bigiel} we have overplotted our star formation rates on
Fig.~8 of \citet{Bigiel2008}, which shows the star formation rate over
the spirals in their sample versus surface density, as well as the
globally averaged star formation rate for the galaxies in the survey
by \citet{Kennicutt1998}. Our simplified model seems to fit the
observational data reasonably well. Our distribution appears to
support the decline in the star formation rate below around
$\Sigma=10$\,M$_{\odot}$\,pc$^{-2}$, as seen in \citet{Bigiel2008}. Our
points also agree with the lower end of those from
\citet{Kennicutt1998}, similarly indicating a possible relative
decline in star formation at low densities. In fact the star formation
rate appears to show a change in slope, and there is much more scatter
at surface densities less than 10\,M$_{\odot}$\,pc$^{-2}$. To clarify
this we have binned the points from our results in
Figure~\ref{number5}. There we show the mean star formation rate at
each surface density, together with scatter indicated as error bars at
one standard deviation. There is evidently more scatter at low
densities and a steeper slope. We discuss this more in the next
sections.

Fig.~\ref{number5} also shows the difference when we take only the
regions with bound gas (blue points), and when all 500 $\times$ 500\,pc 
regions covering the disc are considered, even if they contain no bound gas (red points). For the latter we assume a maximum star formation rate of $10^{-6}$ \,M$_{\odot}$\,kpc$^{-2}$\,yr$^{-1}$ in the squares where there is no bound gas, rather than zero, in order to calculate the error bars. This estimate is approximately the lowest resolvable star formation rate in our simulations.
As expected the slope is
steeper when regions with this minimal star formation rate are
included (red points). Also the figure indicates that regions which do
not contain bound gas tend to have average surface densities of less
than $\sim 10$\,M$_{\odot}$\,pc$^{-2}$.

\subsection{The local Kennicutt-Schmidt law for different tracers}
It has become apparent from recent observations that the star
formation law varies for different tracers. The star formation law is
shallower for high density gas, e.g. H$_2$, HCN
\citep{Gao2004,Bigiel2008}, so $\Sigma_{\rm SFR}$ goes roughly as
$\Sigma(H_2)^{1.0}$. By comparison the star formation rate varies as
approximately $\Sigma^{1.5}$ when $\Sigma$ is the surface density of
all gas, i.e. H\textsc{i} and H$_2$. The law becomes even steeper when
just H\textsc{i} is considered \citep{Heyer2004,K2007,Bigiel2008},
suggesting that H\textsc{i} is not a good measure of local star
formation. \citet{Krum2007b} propose that the transition from superlinear to linear occurs when the density at which a molecule is excited is similar to the median density of the galaxy.

A common feature in the results from the simulations and the results
of \citet{Bigiel2008} is that there is much more scatter at low
surface densities compared to high. A likely explanation is that at
low densities the gas can exhibit a range of distributions -- the gas
can lie in a few dense bound clumps undergoing star formation, or
alternatively in a more diffuse medium with very little star
formation. At higher densities, much more of the gas in a given region
is likely to be in bound clumps undergoing star formation and the
possibility of larger volumes of diffuse gas is
diminished. \citet{Bigiel2008} express this as the filling factor of
the gas, i.e. the ratio of gas as high densities where stars are
forming. The high density tracers preferentially select the top
(highest star formation rate) points from the distribution for all the
gas.

\begin{figure}
\centerline{
\includegraphics[scale=0.48]{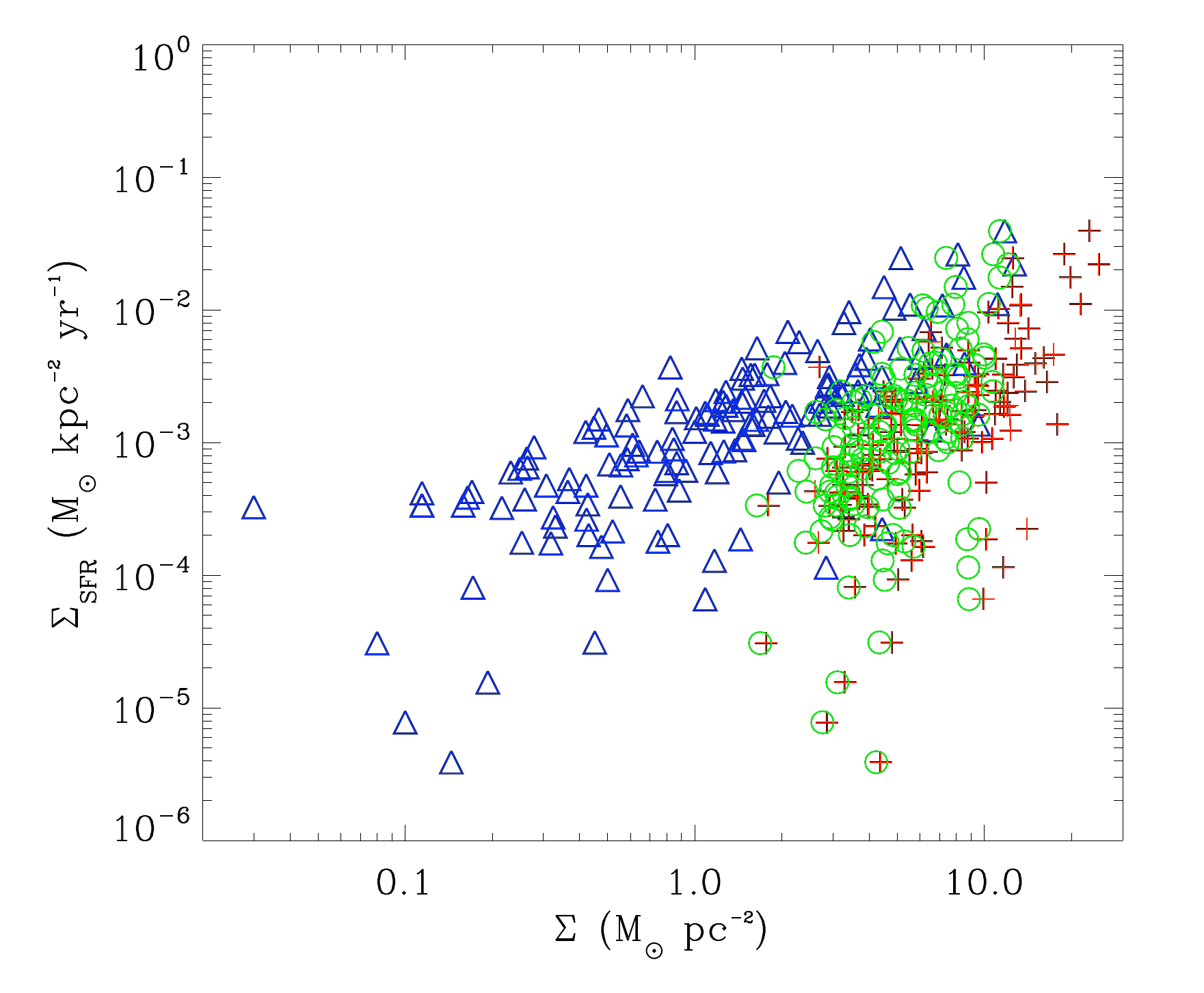}}
\caption{The star formation rate is plotted against surface density
  for 500\,pc$^2$ regions from Model B, where the mean surface density
  is 8\,M$_{\odot}$\,pc$^{-2}$. The red crosses
  represent points where no density cut is imposed. The blue triangles
  correspond to points which only include gas of density above
  10$^{-23}$\,g\,cm$^{-3}$. The slope is clearly shallower when only the
  denser material is used. Finally we selected gas with densities
  $<10^{-23}$\,g\,cm$^{-3}$ (green circles), for which the slope becomes
  steeper.}
\label{number6}
\end{figure}

We can test directly the likely effect of using different tracers in
the simulations. As we have mentioned above the cool component of our
model ISM can be seen as a proxy for molecular gas, whereas the warm
component can be seen as a proxy for atomic gas. It is the cool gas
which is more severely affected by the spiral structure, and therefore
the cool gas which is more likely to be contained in bound
entities. The warm gas, with sound speeds comparable to the potential
depths of the spiral potential, tends to be less affected by the
spiral arms and less easily assimilated into bound clumps. Thus, although
the evolution of different chemical species is not followed in these
simulations, we can use a density cut to select particles over a given
density. In the observations, it is not whether or not the gas is molecular
that defines the slope of the power law, it is merely that the
molecular gas traces the denser parts of the ISM.

Fig.~\ref{number6} shows the star formation rate per unit area against
surface density for Model B when the surface density of all the gas is
used, compared to when there is a density cut of 10$^{-23}$ g
cm$^{-3}$. With the density cut, the surface densities are only
calculated using the gas with density above this threshold, whilst the
star formation rate is the same, thus points are shifted to the left
in the Figure. At low surface densities, much less of the gas is at
high densities, so the points are shifted much further. Hence the
slope is shallower compared to the relation for all the gas and indeed
tends towards the observed linear relation between star formation rate
and surface density for the densest gas. We also selected gas below
this threshold, and as expected a steeper relation ensues. \citet{Gnedin2008} obtain 
similar results by plotting the star formation rate against densities of H\textsc{i}, H$_2$ and H\textsc{i}+H$_2$ calculated in their simulations. 

In our models the slope of $\Sigma_{\rm SFR}$ changes continuously from a very steep slope at low density criterion to linear with a higher density criteria.
The roughly linear relation to surface density for gas above 10$^{-23}$\,g\,cm$^{-3}$ suggests gas above this density is involved in star formation. Actually, only about a quarter of this gas is gravitationally bound. We therefore also calculated $\Sigma_{\rm SFR}$ with a cut of 10$^{-22}$\,g\,cm$^{-3}$, in which case nearly all the gas is bound. The slope is approximately 1.0 in both cases, but the points are shifted to lower surface densities with the higher surface density cut.

\subsection{Theoretical interpretation}

There seems to be evidence from the observations, that for molecular
tracers which presumably correspond to the higher density gas,
$\Sigma_{\rm SFR} \propto \Sigma^n$ with $n=1$.  This relation can be
reproduced in our simulations by considering only the highest density
gas. This relation also holds at the higher surface densities even
when $\Sigma$ is the total surface density. This appears to be
approximately when $\Sigma > 10$\,M$_\odot$\,pc$^{-2}$, where most of
the gas is cool/molecular \citep{DGCK2008,Krum2008}.

This raises two questions: why is the relationship linear, and why is
it much steeper than linear at lower surface densities?

\subsubsection{A shallower local Kennicutt-Schmidt relation?}

At high surface densities a shallower relation of the form
$\Sigma_{\rm SFR} \propto \Sigma^{1.0}$ is found. This contradicts the
straightforward expectation of most theoretical models (see
Section~\ref{introtheory}, and, for example, the discussion in Section
5 of Kennicutt, 1998), in which the star formation rate is presumed to
depend on the local surface density divided by an appropriate
timescale. For example one might take $\Sigma_{\rm SFR} \propto
\Sigma/ t_{\rm dyn} $. Then assuming that $t_{\rm dyn} \propto
\Sigma^{-0.5}$, one finds that $\Sigma_{\rm SFR} \propto
\Sigma^{1.5}$, in rough agreement with the Kennicutt-Schmidt law.

One possible explanation for this discrepancy, as also suggested by \citet{Krum2007b}, might be that star formation only takes place in the densest gas, regardless of the average surface
density. Thus for determining the star formation rate, $\rho$ is
effectively constant, and therefore the local dynamical time given by
a formula such as Equation~\ref{dyntime} is also roughly constant.

\begin{figure}
\centerline{
\includegraphics[scale=0.48]{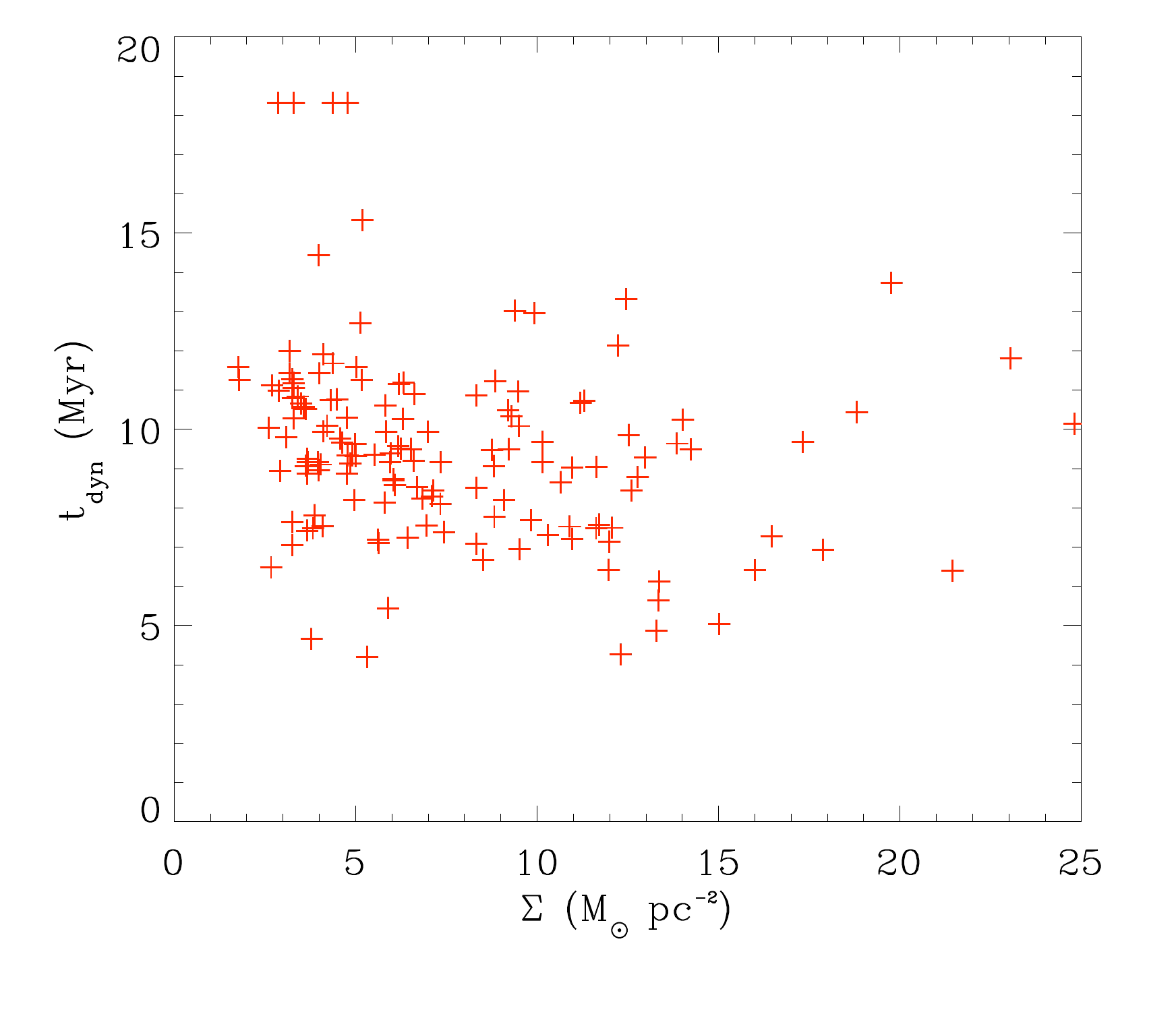}}
\caption{The dynamical time of individual bound entities is plotted
  against the surface density for the 500\,pc$^2$ regions in which they
  lie from Model B, where the mean surface density is 8\,M$_{\odot}$\,pc$^{-2}$.  
  Thus although the dynamical time of each bound clump is related to
  the density of a localised clump of bound gas, this relation largely
  disappears when considering the dynamical times of all the bound gas
  clumps in a 500\,pc$^2$ region.}
\label{number7}
\end{figure}

To examine this further, we plot in Figure~\ref{number7} the various
dynamical times of the various bound clumps against the mean local
surface density of the 500 $\times$ 500\,pc$^2$ areas in which they are
found in Model B, the 8\,M$_{\odot}$\,pc$^{-2}$ simulation. As can be
seen from the figure, the dynamical time shows no particular
correlation with surface density. Thus to a first approximation, the
local star formation rate per unit area just depends in a linear
fashion on the local mean surface density of bound gas. The exception in our calculations is the lowest surface density case (model A) where there is a large increase in the dynamical time at low surface densities.

Our hypothesis differs mainly from \citet{Krum2007b} in that we do not suppose a $\Sigma^{1.5}$ relation for the total gas. Instead, as expressed in Section~3.3, we expect a large degree of scatter between the star formation rate and $\Sigma$ at low surface densities, depending on the local environment of the gas, whilst a linear relation prevails at high surface densities.

\subsubsection{$\Sigma$(total) versus $\Sigma$(H$_2$) }
\label{HImodel}

The difference between the tracers can be illustrated in the following manner.
Supposing that the fraction of the ISM which is in molecular form,
$f(H_2)$, is a monotonically increasing function of the local surface
density $\Sigma$. As an example we take
\begin{equation}
\label{smodel}
f(H_2) = \left( \frac{\Sigma}{\Sigma_{0}} \right)^\alpha,
\end{equation}
for $\Sigma < \Sigma_0$ and for some $\alpha > 0$, and $f=1$
otherwise.  When $\Sigma = \Sigma_0$ let the star formation rate be
$\Sigma_{\rm SFR} = \Sigma_{\rm SFR0}$.

We then assume the observed relation for $H_2$, i.e.  $\Sigma_{\rm
  SFR} \propto \Sigma(H_2)$ to hold at all times, and use this to
calculate the equivalent relation for the total surface density
(H\textsc{i}+H$_2$) gas and for the H\textsc{i} surface density.
Clearly then for $\Sigma > \Sigma_0$, we have the linear relation
$\Sigma_{\rm SFR} \propto \Sigma$.

However for $\Sigma < \Sigma_0$ our assumptions imply that the star
formation law for the total surface density (H\textsc{i} and H$_2$) is
$\Sigma_{\rm SFR} \propto f(H_2) \Sigma \propto \Sigma^{(\alpha +1)}$.
Thus the relation for the total gas is expected to be steeper at low
densities compared to high.

Now consider the relation of star formation rate with surface density
in H\textsc{i}. We then expect there no longer to be a one-to-one
correspondence between star formation rate and surface density! For
example we now have that 
\begin{equation}
\Sigma(HI)=\Sigma-\Sigma(H_2). 
\end{equation}
Then in our simple model (equation~\ref{smodel}) when $\Sigma(HI) =
0$, {\em either} there is no gas at all so that $\Sigma = 0$ and
$\Sigma_{\rm SFR}=0$ {\em or} the gas is fully molecular so that
$\Sigma \ge \Sigma_0$ and $\Sigma_{\rm SFR} \ge \Sigma_{\rm SFR0}$.

Thus overall the star formation rate is multivalued for a given
density of H\textsc{i}. The H\textsc{i} surface density reaches a
maximum at a particular value of $\Sigma$, when the gas starts to
become predominantly molecular. Above this, $\Sigma(HI)$ decreases,
but the star formation rate continues to increase. This is essentially
the top of a parabola-like curve in $\Sigma(HI)$ versus $\Sigma_{\rm
  SFR}$ space, and can be seen in the H\textsc{i} plots of
\citet{Bigiel2008} and \citet{K2007}.

\subsubsection{A change in slope}
For plots of the total surface density, we find a change in the slope around 10\,M$_{\odot}$\,pc$^2$. \citet{Krum2008} interpret this as the surface density at which gas becomes molecular. In our models, this surface density corresponds roughly to the density at which the surface densities of bound and unbound gas are approximately equal. Below this density, the surface density is dominated by low density, unbound gas, and follows a steeper slope, as indicated by the low density criterion on Fig.~6. Above this density, the gas is predominantly bound, thus follows the shallower path. 
Since the gas generally \emph{is} self-gravitating at densities  high enough for molecular gas to be observable \citep{Hart2001}, our critical density is similar to that of \citet{Krum2008}.

\subsubsection{Higher surface densities}
A major disadvantage of the analysis presented in this paper is that,
for computational reasons, we are unable to include very high surface
densities, with no points over $100$~M$_\odot$~pc$^{-2}$. Thus, for
example, unlike previous theories and observations \citep{Kennicutt1989,Krum2005}, we are not
able to claim applicability of our findings to starbursts.

Our simple analysis in Section 3.2.4 suggests that the star formation
rate depends linearly on local surface density, once the gas becomes
fully molecular. Whilst a linear relation is observed for molecular
gas measured using CO in normal galaxies \citep{Bigiel2008}, it is not clear that this
is the case for starbursts. Although \citet{Gao2004} find a linear relation between star
formation rate and luminosity in starbursts for $L(HCN)$, they observe that the star formation
rate varies with CO flux as $L(CO)^{1.4}$. 

Nevertheless, we speculate here that our simple ideas might even be
relevant to the high gas surface densities, provided that for these
HCN is a better indicator of molecular gas mass than CO. Indeed,
various authors (e.g \citealt{Gao2004b,Wu2005,Aalto2008}) have noted that CO
is not a particularly good tracer of star formation in starbursts
compared to HCN. We give two reasons why this might be so.

First, the $^{12}$CO predominantly traces warm molecular envelopes
surrounding cold clouds  \citep{Meier2000,Glenn2001,Gao2004}. The ISM in starbursts is
also considerably more turbulent than that in normal galaxies \citep{Aalto1995}. Thus it is not implausible to suppose that a
substantially smaller fraction of the CO represents self-gravitating
gas. Compared to our interpretation of the \citet{Bigiel2008}
results in Section 3.4.2, it may be that for starbursts we should
regard CO as a proxy for HI and HCN as a proxy for H$_2$.

Second, it seems plausible to assume expect that at high surface
densities optical depth effects start to undermine a simple relation
between L(CO) and L(HCN) and surface density. However, it is the CO
observations which are likely to be effected first. Thus it may be
that at high surface densities, L(CO) underestimates the mass relative
to L(HCN). 

\subsection{Star formation rate versus shock strength}
If, as suggested by \citet{Roberts1969}, star formation is triggered
by spiral density waves, a higher degree of star formation might be
expected in galaxies with stronger shocks. This has been the subject
of much past debate. \citet{Elmegreen1986} argued that since grand
design spirals do not show an increased star formation rate compared
to flocculent galaxies, spiral triggering of star formation is not
significant. Instead the gas and thus the star formation is merely
arranged into spiral arms. Though some recent observations show that
there is a correlation between the SFR and spiral shock strength
\citet{Seigar2002}.

\begin{figure}
\centerline{
\includegraphics[bb=80 350 600 780,scale=0.4]{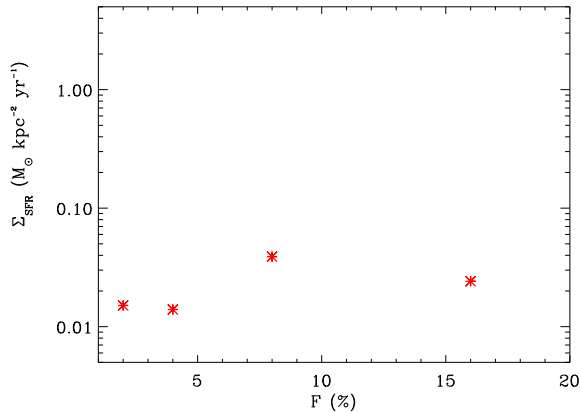}}
\caption{The estimated star formation rate is shown versus the
  strength of the spiral potential when $\Sigma=4$\,M$_{\odot}$\,pc$^{-2}$ 
  (Models A, E, F and G). The star formation remains fairly
  constant as the strength of the potential increases.}
\label{number8}
\end{figure}

For a stronger spiral potential, the stronger shock leads to more gas
in the spiral arms and higher gas densities in the spiral
shock. Therefore we may expect that the amount of bound gas increases
with shock strength. Fig.~\ref{number8} shows the star formation rate
versus the strength of the spiral potential. The star formation rate
does not show an increase with shock strength (within a factor of 2 or
3), but instead remains fairly flat. Thus a stronger shock does not
appear to produce a higher star formation rate (averaged over the
disc).

\begin{figure}
\centerline{
\includegraphics[bb=100 360 600 780,scale=0.43]{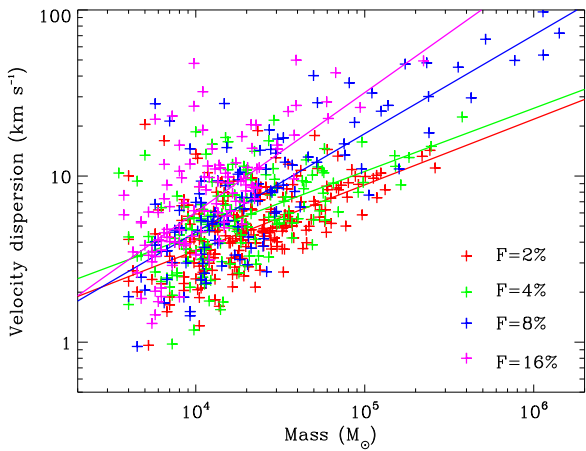}}
\caption{The velocity dispersion is shown against mass for each bound
  clump, from the simulations with different shock strengths. The
  velocity dispersions exhibit higher values at higher shock
  strengths. Best fit lines are also shown through each set of points
  to illustrate that they are offset.}
\label{number9}
\end{figure}

The amount of bound gas depends primarily on the density of the gas
and on the velocity dispersion. Thus a possible explanation for the
apparently constant star formation rate is that the kinetic energy of
the dense gas also increases with the strength of the shock. To test
this, we plot the velocity dispersion of the clumps in
Fig.~\ref{number9}, against the mass of the clumps. The clumps in the
higher shock models clearly have a higher velocity dispersion. Thus
although they have higher densities, there is not a substantially
greater mass of bound gas. However the bound gas is more concentrated
to the spiral arms at higher shock strengths, and therefore there is a
somewhat higher star formation rate \textit{in the spiral arms} at
higher shock strengths (see Section~\ref{interarm}).

\begin{figure}
\centerline{
\includegraphics[bb=100 380 600 750,scale=0.45]{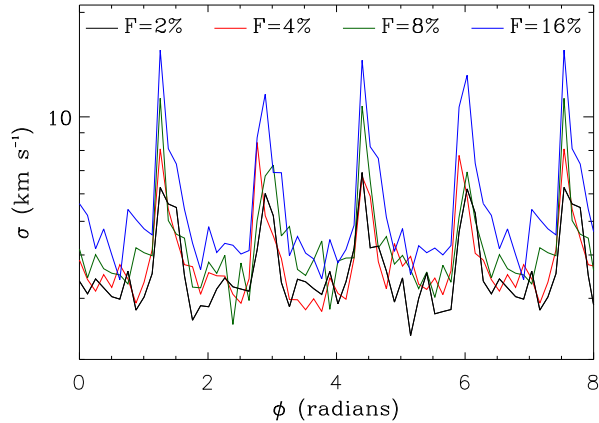}}
\caption{The velocity dispersion is plotted against azimuth for the
  calculation with different shock strengths. The velocity dispersion
  becomes greater in the spiral shock as the strength of the shock
  increases.}
\label{number10}
\end{figure}

We illustrate the velocity dispersion increase further in
Fig.~\ref{number10}, where the velocity dispersion is plotted against
azimuth. The particles used to calculate the dispersion are selected
from a ring of width 200\,pc at a radius of 7.5\,kpc.  The velocity
dispersion of the gas in the spiral arms generally increases as the
shock becomes stronger.

\subsection{Star formation in spiral arm and interarm regions}
\label{interarm}

\begin{figure}
\centerline{
\includegraphics[bb=80 370 600 800,scale=0.45]{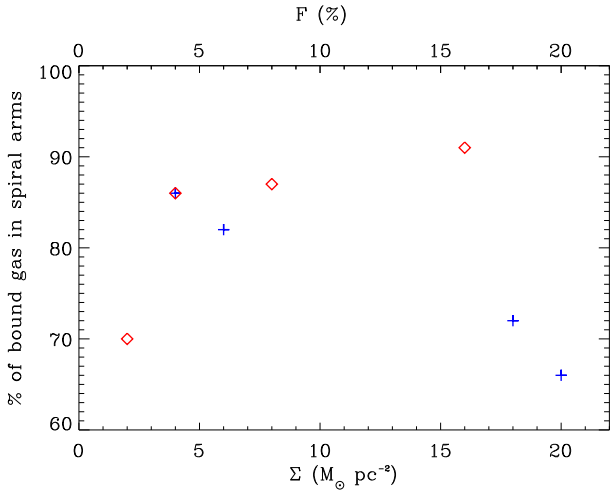}}
\caption{The percentage of bound gas which lies in the spiral arms is
  plotted against surface density (blue crosses) and against strength
  of the potential (red diamonds). At higher surface densities, self
  gravity becomes more important relative to the strength of the
  shock, and less of the bound gas lies in the spiral arms. As the
  strength of the potential increases though (and thus the strength of
  the spiral shock), more of the bound gas lies in the spiral arms.}
\label{number11}
\end{figure}

As described above, the star formation rate, or mass of bound gas,
increases with surface density, but does not vary significantly with
spiral shock strength. Here we investigate whether the degree of star
formation in the spiral arms compared to inter-arm regions varies
according to surface density or shock strength. Gas within a 1\,kpc
wide extent covering the spiral arms is assumed to be spiral arm
material. Figure~\ref{number11} shows the percentage of bound gas in
the spiral arms versus surface density (blue crosses) and the strength
of the potential (red diamonds). In these simulations, between 65 and
90 per cent of the bound gas is located in the spiral arms. The
percentage of bound gas in the spiral arms decreases with surface
density. This indicates that at lower mean surface densities, the
self-gravity of the gas becomes more important compared to the
strength of the shock. Possibly at high surface densities,
gravitational instabilities lead to bound gas in the interarm
regions. However the greatest contribution to the interarm bound gas
is from gas which has become bound in the spiral arms, and remains
mainly bound in the interarm region.
 
The percentage of bound gas which is located in the arms increases
with spiral shock strength. Thus a higher degree of star formation is
likely to occur in the spiral arms for the models with a stronger
spiral potential. Though as discussed in the previous section, the
total star formation rate does not change significantly, instead as
suggested by \citet{Elmegreen1986}, there is more star formation in
the spiral arms simply because more of the gas is there.

The mass of bound gas is calculated using the kinetic, thermal and
magnetic energies. If only the kinetic energy is used, comparatively
more bound gas lies in the interarm regions. However the trends shown
on Fig.~\ref{number11} do not change.

\section{Conclusions}

We have investigated the degree to which the relationship between the
star formation rate per unit area and the surface density of the ISM
in a star-forming galaxy can be understood in terms of simple input
assumptions.

We model the ISM as a self-gravitating, two-phase medium, with one
half the mass at a fixed cool temperature of $T = 100$K, as a proxy
for H$_2$, and the other half at a fixed warm temperature $T = 10^4$K,
as a proxy for H\textsc{i}. We estimate (Section~\ref{SFRestimate})
the local star formation rate in a bound clump as being the mass of
the clump divided by its dynamical timescale, multiplied by an
efficiency factor $\epsilon$ which we take to be $\epsilon = 0.05$ to
give a fit to the observations.

Using these simple input assumptions we find that we can reproduce the
observed relationship which indicates that the local star formation
rate per unit area is linearly proportional to the local surface
density of dense gas (Figure~\ref{number6}). We show that this direct
linear proportionality comes about because (Figure~\ref{number7}) the
local dynamical timescales of bound entities do not correlate with
local mean surface densities.

We also show, in agreement with the observations, that the total
surface density (being a proxy for $\Sigma(HI + H_2)$) and the surface
density of warm gas (being a proxy for $\Sigma(HI)$) are not good
indicators of local star formation rates. A simple model
(Section~\ref{HImodel}) for the dependence of H$_2$ fraction,
$f(H_2)$, as a function of total surface density, in the regime where
$f(H_2) < 1$, provides a simple explanation of why the surface star
formation rate is a steeper function of total surface density in this
regime. Moreover, it is evident from this simple model that the star
formation rate does not have a one-to-one relationship with
H\textsc{i} surface density, implying that any attempt to correlate
star formation rates with H\textsc{i} surface density is likely to
result in large scatter.

We are also able to demonstrate from these simple considerations that
the star formation rate averaged over the galaxy disc does not depend
significantly on the strength of the spiral shock
(Figure~\ref{number8}). This is because, although a stronger shock does
result in a higher gas density downstream of the shock, it also
results in a higher dispersion velocity (Figures~\ref{number9}
and~\ref{number10}). A stronger shock does, however, result in more
of the bound gas, and therefore more of the star formation lying close
to the spiral arms.

\section*{Acknowledgments}
CLD thanks Frank Bigiel for providing an early draft of his paper, and the plot for Fig.~4. We also thank an anonymous referee for suggestions which improved the paper. The calculations reported here were performed using the University of Exeter's SGI Altix ICE 8200 supercomputer.
This work, conducted as part of the award `The formation of stars and planets: Radiation hydrodynamical and magnetohydrodynamical simulations' made under the European Heads of Research Councils and European Science Foundation EURYI (European Young Investigator) Awards scheme, was supported by funds from the Participating Organisations of EURYI and the EC Sixth Framework Programme.  

\bibliographystyle{mn2e}
\bibliography{Dobbs}
\bsp
\label{lastpage}

\end{document}